\documentclass[12pt]{iopart}
\usepackage{iopams}
\usepackage{perpage} %the perpage package
\MakePerPage{footnote} %the perpage package command
\usepackage{graphics}
\usepackage{relsize}
\usepackage{latexsym}
\usepackage{color}
\usepackage{epstopdf}
\usepackage{ctable}
\usepackage{graphicx}
\usepackage{array,multirow,makecell}
\usepackage{verbatim}
\newtheorem{theo}{Theorem}[section]
\newtheorem{uppg}[theo]{Exercise}
\newcommand{\be}{\begin{eqnarray*}}
\newcommand{\ee}{\end{eqnarray*}}
\newcommand{\ben}{\begin{eqnarray}}
\newcommand{\een}{\end{eqnarray}}

\def\subsecn (#1) {\medskip\ \ \ {\it #1}\medskip}

\newcommand{\lp}[1]{\left(\begin{array}{#1}}
\newcommand{\rp}{\end{array}\right)}

\newcommand{\leftd}[1]{\left\{\begin{array}{#1}}
\newcommand{\rightd}{\end{array}\right.}
\setcellgapes{1pt}
\makegapedcells
\newcolumntype{R}[1]{>{\raggedleft\arraybackslash }b{#1}}
\newcolumntype{L}[1]{>{\raggedright\arraybackslash }b{#1}}
\newcolumntype{C}[1]{>{\centering\arraybackslash }b{#1}}

\def\Rb {\mathbb{R}}
\def\Pb {\mathbb{P}}
\def\Eb {\mathbb{E}}

\date{\today}
\begin{document}

\title[]{The three different regimes in coulombic friction}

\author{Azzouz Dermoune$^1$\footnote{Corresponding 
author: Azzouz.Dermoune@univ-lille1.fr}, Daoud Ounaissi$^2$ and Nadji Rahmania$^3$}

\address{$^1$Dermoune Azzouz, Cit\'e scientifique, France}
\address{$^2$Daoud Ounaissi, Cit\'e scientifique, France}
\address{$^3$Nadji Rahmania, Cit\'e scientifique, France}
\eads{\mailto{Azzouz.Dermoune@univ-lille1.fr}, \mailto{daoud.ounaissi@ed.univ-lille1.fr}, \mailto{nadji.rahmania@univ-lille1.fr}}
% For more than one e-mail address, please use the command \eads{\mailto{#1}, \mailto{#2}} with \mailto surrounding each e-mail address.

\begin{abstract}
de Gennes identified three regimes in the phenomenon of the Langevin 
equation wich includes Coulombic friction.
Here we extend and precise this phenomenon to a 
constant external force.
\end{abstract}

\vspace{2pc}

%%%%%%%%%%%%%%%%%%%%%%%%%%%%%%%%%%%%%%%%%%%%%%%

\section{Introduction}
P-G de Gennes \cite{deGennes} studied the Langevin equation under the influence of a dry friction force modelled by the equation 
\be 
dv=-\frac{1}{2}\Delta sgn(v)dt+\sqrt{D}dB,
\ee 
the dry friction force with threshold force $\Delta >0$, and $D >0$ 
is the diffusion coefficient. 
Here $B$ is the standard Brownian motion, and 
$sgn(v)=1$ if $v >0$, and $sgn(v)=-1$ if $v < 0$. 
Comparing the magnitude of $\alpha$, $\Delta$ and $D$ de Gennes \cite{deGennes} identified three different regimes: viscous, partly stuck and stuck. 

Later Touchette et al. \cite{Touchette} extended de Gennes work by calculating 
the time-dependent propagator of the Langevin equation 
\ben 
dv=-\frac{1}{2}[\alpha v-a+\Delta sgn(v)]dt+\sqrt{D}dB, 
\label{touchette} 
\een 
which includes a constant external force $a\in\Rb$.  

In this paper, we precise and extend de Gennes's work to the Langevin equation 
(\ref{touchette}) and find again the result of Touchette et al. \cite{Touchette}
using the trivariate density of Brownian motion, its local and occupation times.  

\section{The three different regimes in coulombic friction} 
If $v(t)$ is solution of (\ref{touchette}), then 
$v(\frac{t}{D})$ satisfies the equation 
\ben 
dv=-\frac{1}{2D}[\alpha v-a+\Delta sgn(v)]dt+dB.
\label{timescale} 
\een 
It follows that for large time $T$ the PDF of the velocity $v(\frac{T}{D})$ 
is approximated by the stationary PDF 
\be 
\frac{1}{Z}\exp\left[-\frac{1}{\nu}\big(
\frac{(v-y)^2}{2\tau}+|v|\big)\right],
\ee 
where 
\be 
Z=\frac{1}{2\nu}
\left[\exp(\frac{\tau-2y}{2\nu})G(\frac{\tau-y}{\sqrt{\tau\nu}})+\exp(\frac{\tau+2y}{2\nu})G(\frac{\tau+y}{\sqrt{\tau\nu}})\right]
\ee 
is the partition function i.e. the normalization constant. Here and the sequel 
\be 
\fl G(u)=\frac{1}{\sqrt{2\pi}}\int_u^{+\infty}\exp(-\frac{v^2}{2})dv,\quad 
\nu=\frac{D}{\Delta},\quad \tau=\frac{\Delta}{\alpha},\quad y=\frac{a}{\alpha}.
\ee 
 
We say that the stochastic process $(V_D: D >0)$ 
defined in some probability space $(\Omega, {\cal F}, \Pb)$ converges
in probability distribution as $D\to 0$ 
to the PDF $f$ if for each couple $l < r$ of real numbers 
\be 
\Pb(l\leq V_D\leq r)\to \int_l^r f(v) dv,\quad\mbox{as}\quad D\to 0. 
\ee   
Now we can announce our result.\\
1) Stuck regime. If $|a| < \Delta$, then the velocity
$v(\frac{T}{D})\to 0$ as $D\to 0$. More precisely 
 $\frac{1}{\nu}v(\frac{T}{D})$
converges in distribution as $D\to 0$ to the PDF 
\be 
\frac{1-y^2}{2}\exp\left[-|v|(1-sgn(v)|y|)\right].
\ee 
Observe that if the constant force $a=0$, then $y=0$ and the limit 
is  
\be 
\frac{1}{2}\exp(-|v|).
\ee 
2) Partly stuck regime. If $|a|=\Delta$, then the velocity
$v(\frac{T}{D})\to 0$ as $D\to 0$. More precisely 
we distinguish two cases. 

a) If we consider only the event $av(\frac{T}{D})< 0$, 
then   
\be 
\frac{1}{\nu} v(\frac{T}{D})\to 2\exp(-2|v|){\bf 1}_{[av < 0]}\quad \mbox{as}\quad 
D\to 0.  
\ee 

b) If we consider only the event $av(\frac{T}{D})> 0$, then
\be 
\frac{1}{\sqrt{\nu}}v(\frac{T}{D})\to \frac{2}{\sqrt{2\pi\tau}}\exp(-\frac{v^2}{2\tau}){\bf 1}_{[av > 0]}\quad \mbox{as}\quad 
D\to 0.
\ee
Moreover the probability of the event $av(\frac{T}{D})> 0$ tends to 1  as $D\to 0$. Hence $\frac{1}{\sqrt{\nu}}v(\frac{T}{D})$ converges to $\frac{2}{\sqrt{2\pi\tau}}\exp(-\frac{v^2}{2\tau}){\bf 1}_{[av > 0]}$.

3) Viscous regime. If $|a|>\Delta$ then as $D\to 0$ the velocity 
$v(\frac{T}{D})$ becomes Gaussian with the mean $(y-sgn(y)\tau)$
and the variance $\nu\tau$. More precisely, we have 
\be 
\frac{v(\frac{T}{D})-(y-sgn(y)\tau)}{\sqrt{\nu}}\to 
\frac{1}{\sqrt{2\pi\tau}}\exp(-\frac{v^2}{2\tau}). 
\ee
\newpage
Observe that the asymptotic mean $y-sgn(y)\tau$ is the minimizer of the potential $v\to \frac{(v-y)^2}{2\tau}+|v|:=U(v)$.  

%%%%%%%%%%%%%%%%%%%%%%%Figure%%%%%%%%%%%%%%%%%%%%%%%%%

\begin{figure}[!ht]
  \centering

     \includegraphics[width=16cm]{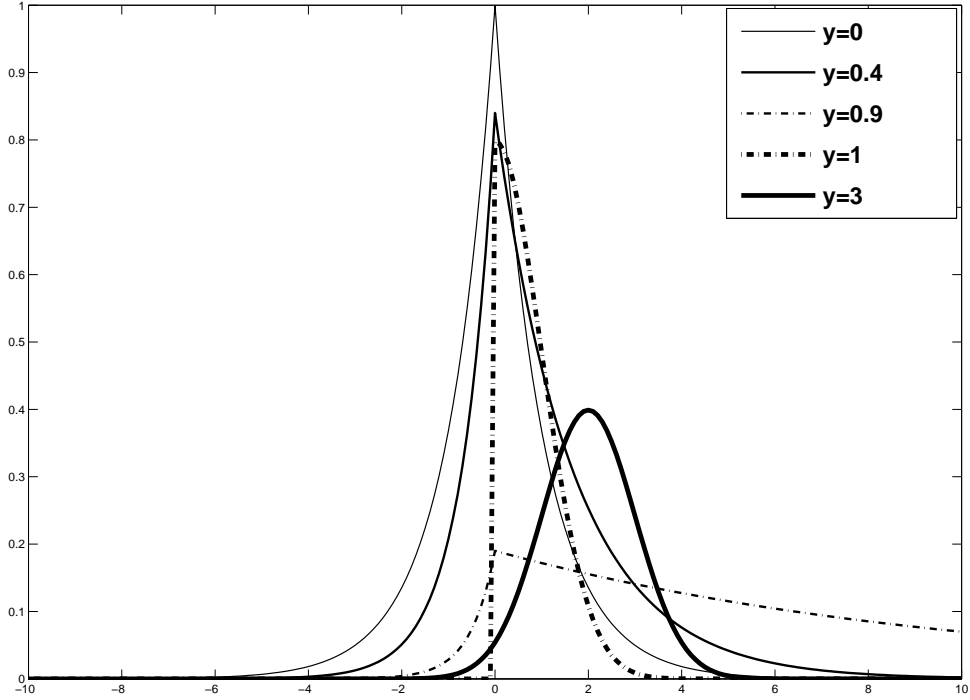}
 \caption{Three scenarios of the stuck regime with $y=0, 0.4, 0.9$, partly stuck regime with $y=1$ and $\tau=1$ and viscous regime with $y=3$ and $\tau=1$.}

\end{figure}

%%%%%%%%%%%%%%%%%%%%%%%%%%%%%%%%%%%%%%%%%%%%%%%%%%%
The proof was done in a general case in \cite{DDN}. For the sake of completeness we 
recall it. It is sufficient to show the case $a\geq 0$ i.e. $y\geq 0$. 
\section{Proof}
\subsection{Stuck regime} 
We observe that the potential $U$ attains its minimum 
$\frac{y^2}{2\tau}$ at $v=0$. We have 
\be 
\Pb(l\leq\frac{v(\frac{T}{D})}{\nu}\leq r)=
\frac{\int_{l\nu}^{r\nu}\exp(-\frac{U(v)}{\nu})dv}
{\int_{-\infty}^{+\infty}\exp(-\frac{U(v)}{\nu})dv}. 
\ee 
Multiplying the denominator and the nominator by $\exp(\frac{y^2}{2\tau\nu})$,
and using the change of variable $\frac{v}{\nu}$ we have   
\be 
\Pb(l\leq\frac{v(\frac{T}{D})}{\nu}\leq r)=
\frac{\int_{l}^{r}\exp\left[-|v|(1-sgn(v)\frac{y}{\tau})-\sqrt{\nu}\frac{v^2}{2\tau}\right]dv}{\int_{-\infty}^{+\infty}
\exp\left[-|v|(1-sgn(v)\frac{y}{\tau})-\sqrt{\nu}\frac{v^2}{2\tau}\right]dv}.
\ee 
The latter converges to 
\be 
\frac{\int_{l}^{r}\exp\left[-|v|(1-sgn(v)\frac{y}{\tau}\right]dv}{\int_{-\infty}^{+\infty}
\exp\left[-|v|(1-sgn(v)\frac{y}{\tau}\right]dv}
\ee 
as $\nu\to 0$, which achieves the proof of the stuck regime. 

\subsection{Partly stuck regime} 
a) We are going to prove for each $l < r\leq 0$ that  
\be 
\Pb(l\leq\frac{v(\frac{T}{D})}{\nu}\leq r\,|\,v(\frac{T}{D})<0)\to 
\int_l^r2\exp(2v)dv. 
\ee 
We have 
\be 
\Pb(l\leq\frac{v(\frac{T}{D})}{\nu}\leq r\,|\, v(\frac{T}{D}) < 0)=
\frac{\int_{l\nu}^{r\nu}\exp\left[-\frac{1}{\nu}(-v+\frac{(v-\tau)^2}{2\tau})\right]dv}{\int_{-\infty}^{0}\exp\left[-\frac{1}{\nu}(|v|+\frac{(v-\tau)^2}{2\tau})\right]dv}. 
\ee 
Multiplying the denominator and the nominator by $\frac{\exp(\frac{\tau}{2\nu})}{\nu}$, and using the change of variable $\frac{v}{\nu}$ we obtain 
\be 
\frac{\int_{l}^{r}\exp\left(2v-\sqrt{\nu}\frac{v^2}{2t})\right)dv}{\int_{-\infty}^{0}
\exp\left(2v-\sqrt{\nu}\frac{v^2}{2t} )\right)dv}.
\ee 
The latter converges to 
\be 
\frac{\int_{l}^{r}\exp(2v)dv}{\int_{-\infty}^{0}\exp(2v)dv}
\ee 
as $\nu\to 0$, which achieves the proof of the part 1. 

b) We have, for $0<l < r$, 
\be 
\Pb(l\leq\frac{v(\frac{T}{D})}{\sqrt{\nu}}\leq r\,|\,v(\frac{T}{D})>0)=
\frac{\int_{l\sqrt{\nu}}^{r\sqrt{\nu}}\exp\left[-\frac{1}{\nu}(v+\frac{(v-\tau)^2}{2\tau})\right]dv}{\int_{0}^{+\infty}
\exp\left[-\frac{1}{\nu}(v+\frac{(v-\tau)^2}{2\tau})\right]dv}. 
\ee
Multiplying the denominator and the nominator by $\frac{\exp(\frac{\tau}{2\nu})}{\sqrt{\nu}}$ and using the change of variable $\frac{v}{\sqrt{\nu}}$ 
we get the proof of the first part of b).\\
For the second part we use the same proof and show that $\Pb(av(\frac{T}{D}) > 0) \to 1$ as $D \to 0$.   
 
\subsection{Viscous regime} 
The main tool of the proof is the following well known result see e.g.\cite{AthreyaHwang}. \\

{\bf Lemma:} \label{Mtool}
 Let $H$ be any measurable map such that 
\be 
\int_{-\infty}^{+\infty} \exp(-H(v))dv < +\infty
\ee 
and 
\be 
inf\{H(v): |v-v_0|\geq \delta\} > H(v_0)
\ee 
for some $v_0$ and $\delta >0$. Then for any $\gamma >0$, 
\be 
\nu^{-\gamma}\int_{[|v-v_0|\geq \delta]}\exp\left[-\frac{1}{\nu}(H(v)-H(v_0))\right]dv\to 0
\ee 
as $\nu\to 0$. 

Now, let us apply this lemma with 
$H(v)=U(v)$ and $v_0=y-\tau$ the minimizer of $U$.  
We have, for $l < r$, 
\be 
\Pb(l\leq\frac{v(\frac{T}{D})-(y-\tau)}{\sqrt{\nu}}\leq r)=
\frac{\int_{l\sqrt{\nu}+y-\tau}^{r\sqrt{\nu}+y-\tau}\exp\left(-\frac{1}{\nu}U(v)\right)dv}{\int_{-\infty}^{+\infty}\exp(-\frac{1}{\nu}U(v)dv}. 
\ee 
We have for $v >0$, that 
\be 
U(v)-U(y-\tau)=\frac{(v-(y-\tau))^2}{2\tau}.
\ee 
If $l >-\infty$, then for small $\nu$, we have  
\be 
\fl \int_{l\sqrt{\nu}+y-\tau}^{r\sqrt{\nu}+y-\tau}\exp\left[-\frac{1}{\nu}(U(v)-U(y-\tau))\right]\frac{dv}{\sqrt{\nu}}
&&=\int_{l\sqrt{\nu}+y-\tau}^{r\sqrt{\nu}+y-\tau}\exp\left[-\frac{1}{2\tau\nu}(v-(y-\tau))^2\right]\frac{dv}{\sqrt{\nu}}\\
&&=\int_{l}^{r}\exp(-\frac{v^2}{2\tau})dv, 
\ee 
and then 
\be 
\Pb(l\leq\frac{v(\frac{T}{D})-(y-\tau)}{\sqrt{\nu}}\leq r)=\int_{l}^{r}\exp(-\frac{v^2}{2\tau})\frac{dv}{\sqrt{2\pi\tau}}.
\ee  

If $l=-\infty$, then 
\be 
\int_{-\infty}^{r\sqrt{\nu}+y-\tau}\exp\left[-\frac{1}{\nu}(U(v)-U(y-\tau))\right]\frac{dv}{\sqrt{\nu}}
=(1)+(2),
\ee 
where 
\be 
(1)&=&\int_{[v <0]}\exp\left[-\frac{1}{\nu}(U(v)-U(y-\tau))\right]\frac{dv}{\sqrt{\nu}}\to 0,\\
(2)&=&\int_{[0\leq v \leq r\sqrt{\nu}+y-\tau]}\exp\left[-\frac{1}{\nu}(U(v)-U(y-\tau))\right]\frac{dv}{\sqrt{\nu}}.   
\ee 
From Lemma (\ref{Mtool}) the term (1) converges to 0. By the change of variable $z=\frac{v-(y-\tau)}{\sqrt{\nu}}$, the term (2) 
\be 
(2)=\int_{[-\frac{(y-\tau)}{\sqrt{\nu}}\leq z \leq r]}\exp(-\frac{z^2}{2\tau})dz
\ee 

converges to $\int_{-\infty}^{r}\exp(-\frac{z^2}{2\tau})dz$. By taking $r=+\infty$, we get 

\be 
\int_{-\infty}^{+\infty}\exp\left[-\frac{1}{\nu}(U(v)-U(y-\tau))\right]\frac{dv}{\sqrt{\nu}}\to \int_{-\infty}^{+\infty}\exp(-\frac{z^2}{2\tau})dz,
\ee 
and then 
\be 
\Pb(l\leq\frac{v(\frac{T}{D})-(y-\tau)}{\sqrt{\nu}}\leq r)\to \int_{l}^{r}\exp(-\frac{v^2}{2\tau})\frac{dv}{\sqrt{2\pi\tau}},
\ee  
which achieves the proof.   
\section{Time-dependent propagator}  
%for $\alpha=a=0$ using local occupation time} 
Now we drop the coefficient $\frac{1}{2}$ in (\ref{touchette})
and we discuss the calculation of the time-dependent propagator 
of 
\be 
dv=-[\alpha v+a+\Delta sgn(v)]dt+\sqrt{D}dB.  
\ee 
Using the equality  of the laws or the probability distributions 
of $(\sqrt{D} B(\frac{t}{D})$ and $(B(t))$, we derive that 
\be 
Law(v^{\alpha,a,\Delta,D}(t))=Law(v^{\frac{\alpha}{D},\frac{a}{D},\frac{\Delta}{D},1}(Dt)).
\ee 
Hence the propagators $p^{\alpha,a,\Delta,D}(v,t\,|\,v_0,0)$, 
$p^{\frac{\alpha}{D},\frac{a}{D},\frac{\Delta}{D},1}(v,t\,|\,v_0,0)$ respectively of 
$v^{\alpha,a,\Delta,D}(t)$ and $v^{\frac{\alpha}{D},\frac{a}{D},\frac{\Delta}{D},1}(t)$ satisfy the relation  
\be 
p^{\alpha,a,\Delta,D}(v,t\,|\,v_0,0)=p^{\frac{\alpha}{D},\frac{a}{\Delta},
\frac{\Delta}{D},1}(v,Dt\,|\,v_0,0).
\ee 
Hence, it sufficient to study the case $D=1$.
\section{Time-dependent propagator for $\alpha=a=0$ using local occupation time} 
We denote by $\Pb$ and $\Pb_{v_0}$ the 
probability distribution respectively of the trajectories $s\in [0,t]\to v(s)$
of the solution of (\ref{touchette}) and the Brownian motion starting from $v_0$. 

Under the probability distribution  
\be 
\exp\left(-\Delta\int_0^t sgn(B_s)dB_s-
\frac{t\Delta^2}{2}\right)d\Pb_{v_0}:=f_{sgn}(B)d\Pb_{v_0}
\ee
the process 
$(B(s): s\in [0,t])$ is solution of the equation
\ben 
dv=-\Delta sgn(v)dt+dB,\quad v(0)=v_0.
\label{00delta}
\een 
We simplify the stochastic integral      
$\int_0^t sgn(B_s)dB_s$ using Tanaka formula \cite{Tanaka}  
\be 
|B_t|=|v_0|+\int_0^t sgn(B_s)dB_s+2L_t.
\ee 
Here the local time  
\be 
L_t=\lim_{\varepsilon\to 0}\frac{1}{4\varepsilon}\int_0^t {\bf 1}_{[|B_s|\leq \varepsilon]}ds\\
=\frac{1}{2}\int_0^t\delta(B_s)ds. 
\ee 
It follows that 
\be 
-\int_0^tsgn(B_s)dB_s=|v_0|-|B_t|+2L_t.  
\ee 
Now, 
\be 
f_{sgn}(B)=\exp\left(\Delta(|v_0|-|B_t|+2L_t)-\frac{t\Delta^2}{2}\right).
\ee 
The densities of $v(t)$ 
and the Brownian motion $B(t)$ are related by  
\be 
p(v,t\,|\,v_0)&=&\Eb_{v_0}\left[\delta(B_t-v)
\exp(\Delta(|v_0|-|B_t|+
2L_t)-\frac{t\Delta^2}{2})\right]. 
\ee 
The latter formula is also known as path integral representation
\cite{Baul}. 
Hence the law of the solution  
$v(t)$ is given by the law of $(B_t, L_t)$. 

\subsection{Density of Brownian motion and its local time}      
Set $\Gamma_t=\int_0^t {\bf 1}{[ B_s\geq 0]}ds$, and  
\be 
h(s,v)=\frac{|v|}{\sqrt{2s^3\pi}}\exp(-\frac{v^2}{2s}),\quad s >0, v\in\Rb.
\ee 
Karatzas and Shreve \cite{Karatzas} have calculated the 
probability density $\Pb_{v_0}(B_t\in db, L_t\in dl, \Gamma_t\in d\tau):=p_t(dv,dl,d\tau\,|\,v_0)$ of $(B_t,L_t,\Gamma_t)$ 
as follows. For $v_0 \geq 0$ we have
\be 
\Pb_{v_0}(B_t\in db, L_t\in dl, \Gamma_t\in d\tau)&=&2h(\tau,l+v_0)h(t-\tau,l-b)dbdl,\quad b < 0,\\
\Pb_{v_0}(B_t\in db, L_t\in dl, \Gamma_t\in d\tau)&=&2h(t-\tau,l)h(\tau,l+b+v_0)dbdl,\quad b >0,\\
\Pb_{v_0}(B_t\in db, L_t=0, \Gamma_t=t)&=&\omega(v_0,b,t),\quad b >0, v_0\geq 0,
\ee 
where 
\be 
\omega(v_0,b,t)&=&\gamma_t(b-v_0)-\gamma_t(b+v_0),\\
\gamma_t(u)&=&\frac{1}{\sqrt{2t\pi}}\exp(-\frac{u^2}{2t}).  
\ee 
We derive the joint distribution of $(B_t, L_t)$ under $\Pb_{v_0}$ with $v_0\geq 0$: 
\be  
\fl \Pb_{v_0}(B_t\in db, L_t\in dl)&=2\frac{(2l+v_0-b)}{\sqrt{2\pi t^3}}
\exp\left[-\frac{(2l+v_0-b)^2}{2t}\right]dbdl,\quad b < 0, l >0,\\
\fl \Pb_{v_0}(B_t\in db, L_t\in dl)&=2\frac{(2l+v_0+b)}{\sqrt{2\pi t^3}}
\exp\left[-\frac{(2l+v_0+b)^2}{2t}\right]dbdl+\omega(v_0,b,t)\delta(l),\quad b > 0, l \geq 0. 
\ee 

Now, we calculate the density of the solution (\ref{00delta}) as follows. 
If $v < 0$, then  
\be 
\fl p(v,t\,|\,v_0)&=\Eb_{v_0}\left[\delta(B(t)-v)\exp\left(\Delta(v_0-|B_t|+2L_t)-\frac{t\Delta^2}{2}\right)\right]\\
&:=\exp\left(\Delta(v_0+v)-\frac{t\Delta^2}{2}\right)
\Eb_{v_0}\left[ \delta(B(t)-v)\exp(2\Delta L_t)\right]\\
&= 2\exp\left(\Delta(v_0+v)-\frac{t\Delta^2}{2}\right)\int_0^{+\infty}\frac{(2l-v+v_0)}{\sqrt{2t^3\pi}}\exp(2\Delta l-\frac{(2l-v+v_0)^2}{2t})dl\\
&=\exp(\Delta(v_0+v)-\frac{t\Delta^2}{2})\int_0^{+\infty}\frac{(l-v+v_0)}{\sqrt{2t^3\pi}}\exp(\Delta l-\frac{(l-v+v_0)^2}{2t})dl\\
&=\exp(\Delta(v_0+v)-\frac{t\Delta^2}{2})\Big[\frac{1}{\sqrt{2t\pi}}\exp(-\frac{(v_0-v)^2}{2t})+\Delta\int_0^{+\infty}\exp(\Delta l-\frac{(l-v+v_0)^2}{2t})\frac{dl}{\sqrt{2t\pi}}\Big]\\
&=\frac{1}{\sqrt{2t\pi}}\exp(-\frac{t\Delta^2}{2})
\exp(\Delta(v_0+v))\exp(-\frac{(v_0-v)^2}{2t})+\\
&\Delta\exp(\Delta(v_0+v)-\frac{t\Delta^2}{2})
\int_0^{+\infty}\exp(\Delta l-\frac{(l+v_0-v)^2}{2t})\frac{dl}{\sqrt{2t\pi}}.
\ee 
After some calculation we obtain 
\be 
\int_0^{+\infty}\exp(\Delta l-\frac{(l+v_0-v)^2}{2t})\frac{dl}{\sqrt{2t\pi}}=\exp(\frac{\Delta^2t}{2})\exp(\Delta(v-v_0))F(\frac{v-v_0+\Delta t}{\sqrt{t}}),
\ee 
where $F(v)=\int_{-\infty}^v \frac{exp(-\frac{u^2}{2})}{\sqrt{2\pi}}du$. Finally for $v_0\geq 0$, $v < 0$, 
we have 
\be 
\fl p(v,t\,|\,v_0)=\left(\exp(-\frac{t\Delta^2}{2})\gamma_t(v_0-v)\exp(\Delta(v_0-v))+
F(\frac{v-v_0+\Delta t}{\sqrt{t}})\right)\Delta\exp(2\Delta v).
\ee 
If $v >0$, then 
\be 
\fl p(v,t\,|\,v_0)&=\Eb_{v_0}\left[ \delta(B(t)-v)\exp\big(\Delta(v_0-|B_t|+2L_t)-\frac{t\Delta^2}{2}\big)\right]\\
&:=\exp\big(\Delta(v_0-v)-\frac{t\Delta^2}{2}\big)
\Eb_{v_0}\Big[ \delta(B(t)-v)\exp(2\Delta L_t)\Big]\\
&=\exp\big(\Delta(v_0-v)-\frac{t\Delta^2}{2}\big)\left[2\int_0^{+\infty}\frac{(2l+v+v_0)}{\sqrt{2t^3\pi}}\exp(2\Delta l-\frac{(2l+v+v_0)^2}{2t})dl+
\omega(v_0,v,t)\right]\\
&=\exp(\Delta(v_0-v)-\frac{t\Delta^2}{2})\omega(v_0,v,t)\\
&+\exp(\Delta(v_0-v)-\frac{t\Delta^2}{2})\int_0^{+\infty}\frac{(l+v+v_0)}{\sqrt{2t^3\pi}}\exp(\Delta l-\frac{(l+v+v_0)^2}{2t})dl\\
&=\exp(\Delta(v_0-v)-\frac{t\Delta^2}{2})\omega(v_0,v,t)+\exp(\Delta(v_0-v)-\frac{t\Delta^2}{2})\Big[\frac{1}{\sqrt{2t\pi}}\exp(-\frac{(v_0+v)^2}{2t})\\
&+\Delta\int_0^{+\infty}\exp(\Delta l-\frac{(l+v+v_0)^2}{2t})\frac{dl}{\sqrt{2t\pi}}\Big]\\
&=\exp(\Delta(v_0-v)-\frac{t\Delta^2}{2})\omega(v_0,v,t)+\frac{1}{\sqrt{2t\pi}}\exp(-\frac{t\Delta^2}{2})\exp(\Delta(v_0-v))\exp(-\frac{(v_0+v)^2}{2t})\\
&+\Delta\exp(\Delta(v_0-v)-\frac{t\Delta^2}{2})
\int_0^{+\infty}\exp(\Delta l-\frac{(l+v_0+v)^2}{2t})\frac{dl}{\sqrt{2t\pi}}.
\ee 
From some calculation we obtain 
\be 
\fl \int_0^{+\infty}\exp(\Delta l-\frac{(l+v_0+v)^2}{2t})\frac{dl}{\sqrt{2t\pi}}=&&\exp(\frac{\Delta^2t}{2})\exp(-\Delta(v+v_0))F(\frac{\Delta t-(v+v_0)}{\sqrt{t}}).
\ee 
Finally if $v >0$, $v_0\geq 0$, then 
\be 
\fl p(v,t\,|\,v_0)&=\exp(\Delta(v_0-v)-\frac{t\Delta^2}{2})\omega(v_0,v,t)+\frac{1}{\sqrt{2t\pi}}\exp(\Delta(v_0-v)-\frac{t\Delta^2}{2})\exp(-\frac{(v_0+v)^2}{2t})\\
&+\Delta\exp(-2\Delta v)F(\frac{\Delta t-(v+v_0)}{\sqrt{t}})\\
&=\frac{1}{\sqrt{2t\pi}}\exp(\Delta(v_0-v)-\frac{t\Delta^2}{2})\exp(-\frac{(v_0-v)^2}{2t})+\Delta\exp(-2\Delta v)F(\frac{\Delta t-(v+v_0)}{\sqrt{t}})\\
&=\left(\exp(\Delta(v_0+v)-\frac{t\Delta^2}{2})\gamma_t(v-v_0)+F(\frac{\Delta t-(v+v_0)}{\sqrt{t}})\right) \Delta\exp(-2\Delta v). 
\ee 

Finally we have for $v, v_0\in\Rb$, that 
\be 
p(v,t\,|\,v_0)=q(v,t\,|\,v_0)\exp(-2\Delta |v|)
\ee 
where 
\be 
q(v,t\,|\,v_0)=\Delta\left(\exp(\Delta(|v_0|+|v|)-\frac{t\Delta^2}{2})\gamma_t(v-v_0)+F(\frac{\Delta t-(|v|+|v_0|)}{\sqrt{t}})\right). 
\ee 
Observe that $q(v,t\,|\,v_0)$ is symmetric, i.e. 
$q(v,t\,|\,v_0)=q^{\Delta}(v_0,t|\,v)$. 
In the language of linear diffusion $m(v)=\exp(-2\Delta |v|)$ 
is the speed measure of the linear diffusion (\ref{00delta}).

\section{The case $a\neq 0$} 
In this case the probability distribution $\Pb$ 
of the solution 
\be 
dv=-[\Delta sgn(v)+a]dt+dB,\quad v(0)=v_0,
\ee 
is also absolutely continuous with respect to $\Pb_{v_0}$ (the probability distribution  
of the Brownian motion starting from $v_0$).  
We have 
\be 
\frac{d\Pb}{d\Pb_{v_0}}(B)=\exp(-\int_0^t(\Delta sgn(B_s)+a)dB_s-\frac{1}{2}\int_0^t(\Delta sgn(B_s)+a)^2ds).
\ee 
After some calculation we have 
\be 
-\int_0^t(\Delta sgn(B_s)+a)dB_s&=&\Delta(|v_0|-|B_t|+a(v_0-B_t))+2\Delta L_t,\\
\int_0^t(\Delta sgn(B_s)+a)^2ds&=&(\Delta^2+a^2)t+2a\Delta(2\Gamma_t-t). 
\ee 
It follows that 
\be 
\fl p(v,t\,|\,v_0)=\exp\left[\Delta(|v_0|-|v|+a(v_0-v))-\frac{(\Delta-a)^2t}{2}\right]\Eb_{v_0}\left[\delta(B_t-v)\exp(2\Delta L_t-2a\Delta\Gamma_t)\right]. 
\ee 
Then $p(v,t\,|\,v_0)$ is calculated using 
the trivariate probability distribution $p_t(db,dl,d\tau)$ of $(B_t,L_t,\Gamma_t)$ 
as follows: 
\be 
\fl p(v,t\,|\,v_0)=\exp\left[\Delta(|v_0|-|v|+a(v_0-v))-\frac{(\Delta-a)^2t}{2}\right]\int_0^{+\infty}\int_0^t\exp(2\Delta( l-a\tau)) p_t(v,dl,d\tau). 
\ee 
\section{The general case}
Similarly as above the density of the solution of 
\be 
dv=-[\alpha v+\Delta sgn(v)+a]dt+dB,\quad v(0)=v_0,
\ee 
is 
\be 
\fl p(v,t\,|\,v_0)=
\exp\left[\Delta(|v_0|-|v|+a(v_0-v))-\frac{(\Delta-a)^2t}{2}+\frac{\alpha t}{2}\right]\\ 
\fl \int_0^{+\infty}\int_0^t\int_{0}^{+\infty}\int_{0}^{+\infty}\int_{-\infty}^{+\infty}
\exp\big(2\Delta l-2a\Delta\tau-\frac{\alpha^2}{2}b_2-\alpha\Delta |b_1|-a\alpha b_1\big)p_t(v,dl,d\tau,db_1,d|b_1|,db_2),
\ee 

where $p_t(db,dl,d\tau,db_1,d|b_1|,db_2)$ is the probability density of 
\be 
(B_t,L_t,\Gamma_t,\int_0^tB_sds,\int_0^t|B_s|ds,\int_0^tB_s^2ds).
\ee  

\section{Conclusion} 
We have precised and extended the three different regimes of the Langevin 
equation which includes a viscous friction force, 
a Coulombic friction and a constant external force. Moreover we find again 
its time-dependent propagator using the density of Brownian motion, its local and occupation times.

\end{document}